\documentclass[amsmath,amssymb,showpacs,floatfix,aps,prl,twocolumn,superscriptaddress]{revtex4-1}
\usepackage{graphics}
\usepackage{dcolumn}

\begin{document}

\title{Optoelectronic Properties and Excitons in Hybridized \\Boron Nitride and Graphene Hexagonal Monolayers}
\author{Marco Bernardi}
\affiliation{%
Department of Materials Science and Engineering, Massachusetts Institute of Technology, \\
77 Massachusetts Avenue, Cambridge MA 02139-4307, USA}
\author{Maurizia Palummo}
\affiliation{%
Dipartimento di Fisica, Universit$\grave{a}$ di Roma Tor Vergata, NAST, and European Theoretical Spectroscopy Facility (ETSF), Via della Ricerca Scientifica 1, 00133 Roma, Italy}

\author{Jeffrey C. Grossman}
\email[E-mail: ]{jcg@mit.edu}
\affiliation{%
Department of Materials Science and Engineering, Massachusetts Institute of Technology, \\
77 Massachusetts Avenue, Cambridge MA 02139-4307, USA}

\date{\today}

\begin{abstract}
We explain the nature of the electronic band gap and optical absorption spectrum of Carbon - Boron Nitride (CBN) hybridized monolayers using density functional theory (DFT), GW and Bethe-Salpeter equation calculations. The CBN optoelectronic properties result from the overall monolayer bandstructure, whose quasiparticle states are controlled by the C domain size and lie at separate energy for C and BN without significant mixing at the band edge, as confirmed by the presence of strongly bound bright exciton states localized within the C domains. The resulting absorption spectra show two marked peaks whose energy and relative intensity vary with composition in agreement with the experiment, with large compensating quasiparticle and excitonic corrections compared to DFT calculations. The band gap and the optical absorption are not regulated by the monolayer composition as customary for bulk semiconductor alloys and cannot be understood as a superposition of the properties of bulk-like C and BN domains as recent experiments suggested. 
\end{abstract}


\maketitle

\section*{}
\begin{figure}
\includegraphics{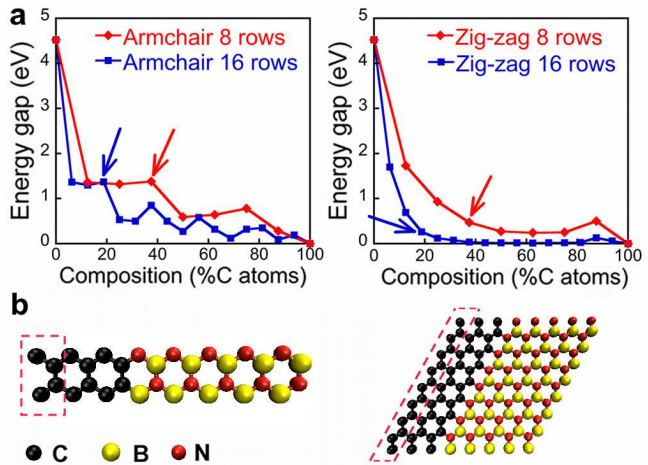}
\caption{(color online) (a) Kohn-Sham band gap calculated using DFT for CBN superlattices with armchair (left) and zig-zag (right) edge. Starting from pure BN (zero \%C), each consecutive point for increasing C concentrations corresponds to the addition of a C row to the structure. (b) Unit cells employed in the 8-rows calculations for the armchair (left) and zig-zag (right) cases. The parts referred to as "one row" in the text are shown in dashed boxes. \label{fig1}}
\end{figure}

\begin{figure}
\includegraphics{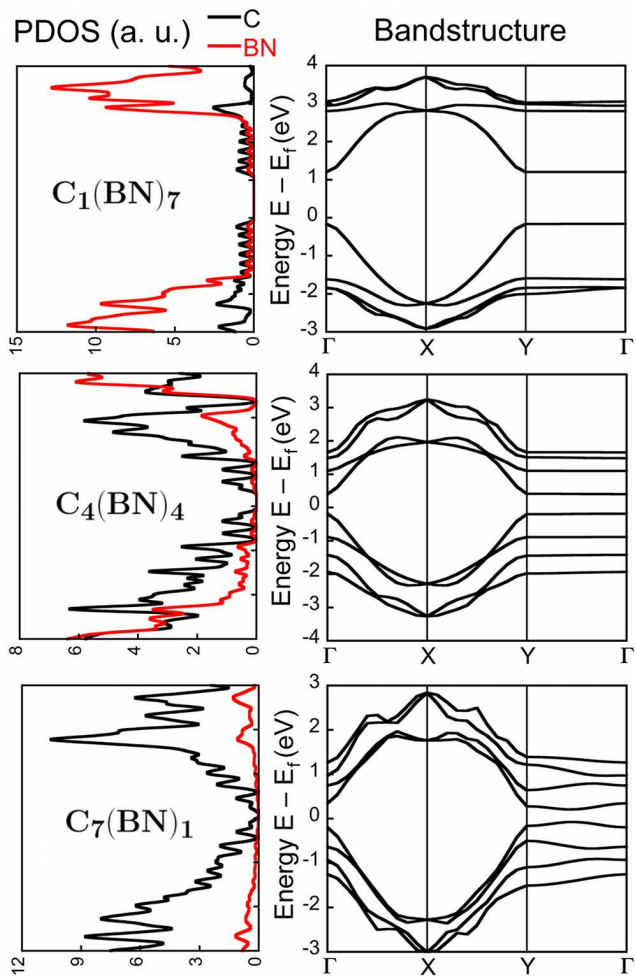}
\caption{(color online) PDOS and corresponding bandstructure plots for three armchair superlattices. When the number of C rows is increased, the band gap decreases continuously. However, the Dirac cone typical of graphene only starts forming for $1-2$ residual BN rows. The PDOS shows that the states close in energy to the gap are mainly due to mixing of C valence states with almost no contribution from BN.  \label{fig2}}
\end{figure}

Recently, two-dimensional (2D) materials with a finite band gap and high carrier mobility have been synthesized and characterized, chiefly in view of their use in optoelectronic devices \cite{Aja, Mos2}. Among these, monolayers of hybridized carbon and boron nitride (CBN) present highly appealing optoelectronic properties due to their tunable optical gap, physically deriving from the large band gap difference between pure hexagonal BN (\textit{h}-BN, energy gap $E_{g} > 5.0$ eV) and graphene, a semimetal with zero energy gap \cite{Graphene}. The C and BN phases are immiscible in 2D, leading to phase separation in the sheet with formation of distinct C and BN domains \cite{Aja}.\\
\indent
The optical absorption spectrum of CBN measured in the recent work from Li \textit{et al.} \cite{Aja} shows two main absorption edges located (using Tauc's extrapolation procedure \cite{Tauc}) at around 1.6 eV and 4.5 eV for a sample with 65\% C content, while for a higher C concentration such absorption peaks were shifted to lower energies. These two features were attributed to the existence of C and \textit{h}-BN domains large enough to resume their individual band gap identities, rather than behave as a CBN alloy.\\
\indent
The electronic band gap of the CBN monolayer has been calculated for different C and BN domain geometries including quantum dot and superlattices that approximate the experimental structure, using tight-binding and DFT calculations \cite{Armchair,CBN2,CBN3,CBNdots, extra1, extra2, extra3}. \\
\indent
For quantum dot geometries, the band gap tunability arises due to confinement in both dimensions \cite{CBNdots}, but DFT level calculations were only applied to limited domain sizes up to $\approx $1 nm, well below the experimental value. CBN 2D superlattices can provide band gap values closer to the experimental regime due to lack of confinement in both dimensions. However, the results appeared in the literature up to now have addressed the band gap tunability as a function of \textit{composition} \cite{CBN2} rather than domain size, or the differences between hydrogenated C nanoribbons and C nanoribbons formed within a CBN superlattice \cite{Armchair,GNR}, without attempting to interpret the experimental optical absorption spectrum in Ref. \cite{Aja} and its connection to the calculated bandstructure. \\
\indent
Here we shed light on the optoelectronic properties of CBN monolayers and show that the double optical peak measured in \cite{Aja} is inherently related to the bandstructure of the CBN system, regardless of the composition and domain size. Such excitation peaks are not due to large C and BN domains, but rather to the near-absence of mixing of C and BN states at the valence band maximum (VBM) and conduction band minimum (CBM), generating two distinct sets of optical transitions: low energy transitions between C states near the band gap resulting in bound bright excitons localized within the C domains, and higher energy transitions from states that are mostly BN-like in character and lie deeper within the conduction and valence bands.\\
\indent
In addition, we demonstrate that composition is not the main variable regulating the CBN electronic structure, band gap and optical absorption, in contrast with the well-known behavior of bulk semiconductor alloys whose energy gap varies continuously with the concentration of the composing elements \cite{Alloy}. We also find large corrections to the DFT band gap using GW calculations and strongly bound excitons, both effects contributing to an optical absorption spectrum similar to the random-phase approximation (RPA) DFT spectrum due to large error compensations in the latter. \\
\indent
We carried out \textit{ab-initio} DFT calculations using the Quantum Espresso code \cite{QE} separately on CBN superlattices with armchair and zig-zag edges, with a total number of C and BN dimer rows of 8 and 16 in distinct sets of calculations, so that electronic structure data are available for a given composition for two different C and BN domain sizes. The Perdew-Burke-Ernzerhof exchange-correlation functional \cite{PBE} is adopted and ultrasoft psuedopotentials \cite{USPP} are used to describe the core electrons. An orthorhombic unit cell with 32 atoms was used for the armchair structures and a hexagonal unit cell with 128 atoms was used for the zig-zag structures (Fig. \ref{fig1}), with interlayer spacing of 15 \AA  \hspace{1pt} in both cases. A kinetic energy cutoff of 35 Ry was used for the plane-wave basis set and of 200 Ry for the charge density.  Converged Monkhorst-Pack \textit{k}-grids \cite{Kgrid} of $3\times3\times1$ and $15\times5\times1$ were used for the zig-zag and armchair cases, respectively.\\
\indent
The GW and Bethe-Salpeter equation (BSE) calculations \cite{MBPT} were performed on three armchair cases 
\footnote{We calculated the formation energy of the armchair and zig-zag type edges and inferred a ratio of 2.5 between the thermal equilibrium length of the armchair and zig-zag domain edges at intermediate compositions and room temperature. This justifies our more accurate study of the optical properties and band gap for the armchair edge type as it is expected to contribute more significantly to the experimental results observed in \cite{Aja}}
with different C concentrations using the Yambo code \cite{Yambo}
\footnote{For the calculations performed with the Yambo code, the ground state Kohn-Sham wavefunctions and eigenvalues were obtained using an LDA exchange-correlation functional (Phys. Rev. B 23, 5048, 1981) with Troullier-Martins norm-conserving pseudopotentials (Phys. Rev. B 43, 1993, 1991) as implemented in the Quantum Espresso code. A kinetic energy cutoff of 55 Ry was used for the wavefunction. These calculations yielded identical results to the ones using the PBE exchange-correlation functional with ultrasoft pseudopotentials, and were carried out with the only scope of obtaining an input compatible with the Yambo code.}
. Briefly, a plasmon-pole model was adopted for the self-energy and cut-off energies of 35 Ry and 5 Ry were used, respectively, for the exchange and correlation part of the self-energy; the Coulomb interaction was truncated in the direction perpendicular to the sheet to avoid spurious interaction with the image system. The GW calculations were performed without self-consistency in the Green's function and the screened Coulomb interaction ($\mathrm{G_{0}W_{0}}$ approximation scheme). A total number of 300 bands ($> 150$ empty bands) was used, together with a converged \textit{k}-point grid of $16\times6\times1$. Both DFT and GW levels of theory were employed in combination with RPA or BSE calculations to compute the optical absorption spectra.\\
\indent
Fig. \ref{fig1}(a) shows the evolution of the DFT band gap as a function of C concentration for armchair superlattices with a simulation cell constituted by a total of 8 and 16 atom rows. For a given composition, the two cases with different number of atom rows give different band gap values; a same band gap value for the two systems is found when comparing superlattices with a same number of C rows (and thus a same C domain size). For example, the arrows in Fig. \ref{fig1}(a) point at structures with three C rows, corresponding to different concentrations for systems with a total of 8 and 16 rows in the unit cell, and yet showing a same energy gap to within 0.1 eV. This behavior is observed for all such pairs of structures with a same number of C rows.\\ 
\indent
A similar trend is found for the zig-zag case (Fig. \ref{fig1}(a)) where systems with a given number of C rows, but corresponding to different C concentrations, show a same energy gap to within 0.1 eV. As observed in previous work \cite{Armchair}, the addition of a single C row causes a drop in the band gap from 4.6 eV for pure \textit{h}-BN to $< 2.0$ eV within DFT calculations.\\
\indent
A more detailed analysis of the DFT electronic structure is presented in Fig. \ref{fig2}, where the CBN systems are labelled $\mathrm{C_{x}(BN)_{(8-x)}}$ for superlattices with x carbon rows out of a total of 8 rows. A direct band gap was found for all the armchair and zig-zag systems studied, thus justifying the high optical absorption observed experimentally. In the armchair cases shown here, the band gap closes progressively for increasing C concentrations but the formation of the Dirac cone only occurs when $1-2$ residual BN rows are present, as seen by the increasing dispersion in the $Y-\Gamma$ direction (normal to the domain edge) resulting in the closure of the gap at the equivalent of the $K$ point of the hexagonal lattice
\footnote{See Ref. \cite{CBN2} for the nomenclature of the Brillouin zone for the orthorhombic cell used here. The $K$ point of the hexagonal lattice, where the graphene band gap is expected to close at high C concentration, falls between the $\Gamma$ and $Y$ point within this scheme.}
. We interpret this behavior as a sign of the incipient delocalization of the VBM and CBM states when the system is close to being a sheet of pure graphene.\\
\indent
The projected density of states (PDOS) shows that for all the compositions studied the states with energy near the gap are mainly due to C, while states farther in energy from the gap result from a clear hybridization of C and BN states. The zig-zag case shows analogous behavior (see Supplemental Material
\footnote{See Supplemental Material at URL for the bandstructure and DFT-RPA optical absorption spectra for the zig-zag CBN system, and for supplemental evidence of the dependence of the band gap on C domain size rather than composition.}
), with subtle differences in the band gap closing near the pure C composition that occurs through a band-crossing mechanism for the formation of the Dirac cone at the $K$ point, causing the slight bump seen in the band gap values in Fig. \ref{fig1}(a) at large C fractions.\\
\indent
The dependence of the band gap on the C domain size, rather than on the overall CBN layer composition [20], can be regarded as an electronic size effect, whereby the C states near the band gap are confined by an effective quasiparticle barrier within the conduction and valence bands formed by the BN states, as seen by the analysis of the PDOS. Nevertheless, the detail of the edge type and binding at the CÐBN interface is also relevant to determine the specific value of the gap (and consequently of the optical properties), as shown by the different band gap values for the zig-zag and armchair cases.\\
\indent
Next, we analyze the results from the GW and BSE calculations.
For the three armchair cases studied with beyond-DFT methods, we found quasiparticle (GW) band gap values significantly higher than the DFT gap (Table \ref{table1}), with corrections in the range of $0.8 - 1.8$ eV. On the other hand, the optical gap (GW-BSE) values were found to be within 0.4 eV of the DFT gaps due to large exciton binding energies in the range of $0.7 - 1.5$ eV (Table \ref{table1}), inferred from the difference between the quasiparticle (GW) and optical (GW-BSE) gaps.\\ 
\noindent
\begin{table}[htbp] 
\caption{Values of the band gap (in eV) within different approximations for three armchair structures studied with beyond-DFT methods. The exciton binding energy (in eV) is also shown. \label{table1}}
\begin{ruledtabular}
\begin{tabular}{ c c c c c }
\vspace{1mm}
Structure&
DFT &
GW&
GW+BSE &
Exciton b. e.\\
\colrule
\vspace{1mm}
$\mathrm{C_{1}(BN)_{7}}$ (12.5\% C)	&	1.39	&	3.25		& 1.80 	&	1.45 \\
\vspace{1mm}
$\mathrm{C_{4}(BN)_{4}}$ (50\% C)	&	0.60	&	1.72		& 0.59 	&	1.04 \\
\vspace{1mm}
$\mathrm{C_{7}(BN)_{1}}$ (87.5\% C)	&	0.30	&	1.04		& 0.32 	&	0.72 \\
\end{tabular}
\end{ruledtabular}
\end{table}
The DFT-RPA absorption spectra for the armchair (Fig. \ref{fig3}(a)) and zig-zag (see Supplemental Material) cases show two absorption peaks resulting from the DFT bandstructures in Fig. \ref{fig2}, with transitions due to the direct band gap along the $Y-\Gamma$ direction (low energy absorption onset) and to the direct gap from higher energy states formed at the $X$ point. These two features are retained in the GW-BSE spectrum (Fig. \ref{fig3}(b)) and can ultimately be indicated as the reason for the two absorption peaks observed experimentally in \cite{Aja}.\\ 
\indent
Nevertheless, the GW-BSE spectrum contains information about two important physical effects: the opening of the quasiparticle gap due to electron-electron interaction at the GW level of theory, causing by itself an almost rigid blue-shift of the absorption spectrum (GW-RPA curve in Fig. \ref{fig3}(b)), and the formation of bound excitonic states resulting in a red-shift of the spectrum back to energies similar to the DFT-RPA approximation level, as seen in the GW-BSE spectra in Fig. \ref{fig3}(b). This explains the excellent qualitative agreement of the DFT-RPA optical absorption spectrum seen in Fig. \ref{fig3}(b), due to a compensation of large errors within DFT.\\ 
\indent
\begin{figure*}[!t]
\includegraphics{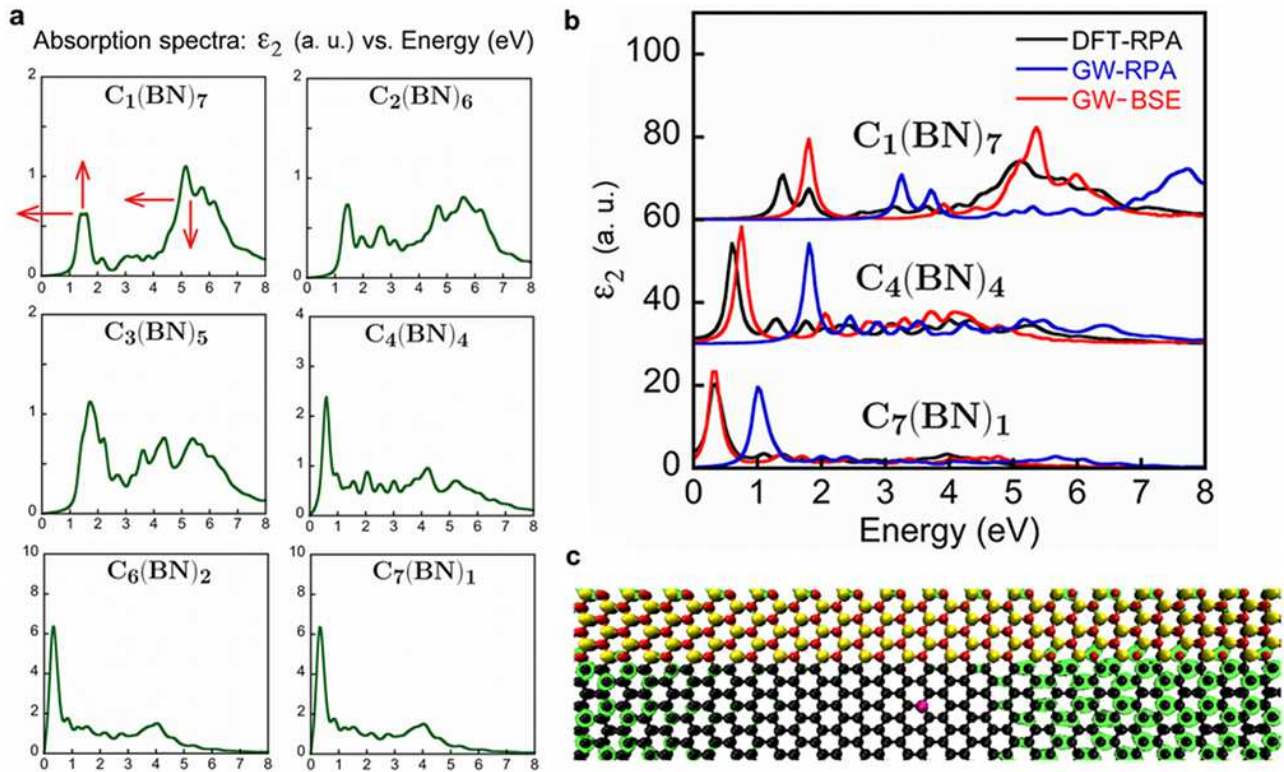}
\caption{(color online) (a) Evolution of the DFT-RPA optical absorption (expressed as the imaginary part of the dielectric tensor, $\epsilon_2$) for increasing sizes of the C domain, for the armchair structures with 8 atom rows shown in Fig. \ref{fig1}(a). The two absorption peaks red-shift and vary their relative strength for an increasing C domain size as pointed by the red arrows. (b) Comparison of absorption spectra for the armchair structures discussed in Table \ref{table1}. The imaginary part of the dielectric tensor (averaged for the three directions) is shown for different levels of approximation including DFT-RPA, GW-RPA and GW-BSE. (c) Exciton spatial distribution (shown in green) for the $\mathrm{C_{4}(BN)_{4}}$ armchair superlattice when the hole is fixed in the position represented by the magenta spot in the C domain. The atoms coloring follows the legend in Fig. \ref{fig1}(b). \label{fig3}}
\end{figure*}
Consistent with the experiments in \cite{Aja}, for increasing C concentrations both absorption peaks are red-shifted and their relative intensity varies, thus showing the tunability of the optical properties of the CBN sheet. By comparing with Ref. \cite{Aja} where systems with over 65\% C concentration show an optical gap of 1.5 eV, we infer that such a low energy absorption edge arises from small C domains with at least one dimension of $1-2$ nm length. However, we note that the Tauc's extrapolation procedure \cite{Tauc} used in Ref. \cite{Aja} can lead to significant errors in the estimation of the absorption onset, and suggest that the experimental spectrum should be regarded as formed by the superposition of the spectra shown in Fig. \ref{fig3} weighed for the different C domain sizes present in the system, a fact justified by the absence of a well-defined low energy peak in the experimental absorption data \cite{Aja}.\\
\indent
The exciton wavefunction obtained within the BSE framework (Fig. \ref{fig3}(c)) is found to be confined by an effective quasiparticle gap at the CBN interfaces: when the hole is fixed in a given position within the C domain, the electron wavefunction localizes within the same C domain, compatible with the presence of VBM and CBM states with C character, and BN-like states farther in energy from the gap. Such delocalized, yet strongly bound bright excitons are a unique feature of nanoscale systems that has been predicted previously for 1D armchair graphene nanoribbons \cite{Zhu}.\\
\indent
In summary, we elucidate the mechanisms underlying the band gap formation and optical processes including absorption and excitons in the CBN monolayer. We find that the energy gap is uniquely regulated by the size of the C domains regardless of the overall system composition, a novel mechanism in clear contrast with the behavior of bulk semiconductor alloys. The absorption spectra of CBN monolayers show two main absorption peaks whose strength and position depend on the C domain size, with transitions between the VBM and CBM states contributing to the low energy absorption onset and to the formation of strongly bound excitons within the C domains. Large quasiparticle and excitonic corrections show that DFT calculations are inadequate to predict the experimental behavior of 2D CBN alloys. The unique dependence of the band gap and absorption spectrum on domain size rather than concentration opens new possibilities for band gap engineering and in general for optoelectronic and photovoltaic applications distinct from those existing for \textit{h}-BN and graphene. Similar large excitonic effects have been predicted recently for $\mathrm{MoS_{2}}$ monolayers \cite{Mos2exc}, suggesting that this new family of 2D nanomaterials could constitute a novel playground for optoelectronic and excitonic devices.
\begin{acknowledgments}
M.B. acknowledges funding from Intel through the Intel Ph.D. Fellowship. We wish to thank NERSC for providing computational resources.
\end{acknowledgments}
\bibliography{references}

\end{document}